# Inferring Social Structure and Dominance Relationships Between Rhesus macaques using RFID Tracking Data

HANUMA TEJA MADDALI, MICHAEL NOVITZKY, BRIAN HROLENOK, DANIEL WALKER, and TUCKER BALCH, Georgia Institute of Technology, Atlanta
KIM WALLEN, Emory University, Atlanta

## 1. INTRODUCTION

In this paper we address the problem of inferring social structure and dominance relationships in a group of rhesus macaques (a species of monkey) using only position data captured using RFID tags. Automatic inference of the social structure in an animal group enables a number of important capabilities, including: 1) A verifiable measure of how the social structure is affected by an intervention such as a change in the environment, or the introduction of another animal, and 2) A potentially significant reduction in person hours normally used for assessing these changes. Social structure in a group is an important indicator of its members' relative level of access to resources and has interesting implications for an individual's health [Tunga et al. 2012] and learning in groups [Drea and Wallen 1999]. There are two main quantitative criteria assessed in order to infer the social structure; Time spent close to conspecifics, and displacements. An interaction matrix is used to represent the total duration of events detected as grooming behavior between any two monkeys. This forms an undirected tie-strength (closeness of relationships) graph. A directed graph of hierarchy is constructed by using the well cited assumption of a linear hierarchy for rhesus macaques [Brent et al. 2013][Maestripieri and Hoffman 2012]. Events that contribute to the adjacency matrix for this graph are withdrawals or displacements where a lower ranked monkey moves away from a higher ranked monkey. Displacements are one of the observable behaviors that can act as a strong indication of tie-strength and dominance. To quantify the directedness of interaction during these events we construct histograms of the dot products of motion orientation and relative position. This gives us a measure of how much time a monkey spends in moving towards or away from other group members. We present results of an analysis based on these approaches for a group of 6 male monkeys that were tracked over a period of 60 days at the Yerkes National Primate Research Center[1].

## 2. APPROACH

The experimental setup is the same as in [Huang et al. 2012]. The monkeys reside in a 3m x 3m metal enclosure. An Ubisense real-time locating system provides tag-id, timestamp and 3D position tuple outputs. This data is aggregated for each of the 4 tags on the collars worn by the monkeys. The reading of these tags are filtered and averaged to get a position estimate of each monkey.

---

[1]This research was supported by the National Institute for Mental Health (MH050268-14S1) as well as by the National Center for Research Resources to the Yerkes National Research Center (P51 RR00165; YNRC Base grant), which is currently supported by the Office of Research Infrastructure Programs/OD P51OD11132. The YNPRC is fully accredited by Americans for the Assessment and Accreditation of Laboratory Care, International.

## 2.1 Inferring tie-strength

A necessary prerequisite for most types of interaction between individuals is spatial proximity. Individuals cannot cooperate if they are not close enough to perceive that their assistance is needed or desired and to provide said service within an appropriate time frame [Crofoot et al. 2011]. In our current work we measure tie-strength by detecting events of passive interaction. Tie-strength for monkeys i and j is the time spent per day engaging in events such as grooming or passive interaction. The male macaques in our dataset have grooming sessions that last roughly 1 minute. Grooming events are therefore characterized by 2 stationary monkeys within a threshold distance of a macaque's arms length (0.5m) and a duration of at least 1 minute. This data gives us an affiliation matrix $A$ for the social network of the monkeys where each element $[A]_{ij}$ of the matrix represents the tie-strength between monkeys $i$ and $j$. An estimate of the sociability of an individual can be derived from the weighted degree of each node. The greater the weighted degree, the more that monkey interacts with the others in the group. Figure 1 shows the Affiliation graph representing matrix $A$. We see that Monkey 6 and Monkey 2 are an example of a pair that does not interact frequently. This can also be visualized using Heat Maps as shown in Figure 3. We can clearly see that the region of the enclosure frequented by Monkey 2 (indicated by bright yellow) does not overlap with that of Monkeys 5 or 6. Whereas Monkeys 5 and 6 have strongly overlapping regions and, from Figure 1, strong affiliation.

## 2.2 Inferring dominance relations

After dominance relations have been established between the agents a subordinate individual will usually avoid the dominant one or express fear and submission in his presence [Maestripieri and Hoffman 2012]. Interactive behaviors between monkeys can be inferred by detecting individual behaviors that act as strong indicators of dominance. Some behaviors that can be ascertained using position and velocity data include withdrawals, displacements, attacking and chasing. For example, if two monkeys are detected to be running while within proximity of each other there is a high probability this is a chasing behavior. Dominance information may then be extracted from all the interactive behaviors detected as chasing. Withdrawal is characterized by a lower ranking individual A changing its trajectory with a slightly higher exit velocity to allow a higher ranking individual B to continue unimpeded. Displacement is similar to withdrawal except that the lower ranking individual A is initially stationary, for example, near a feeding area. The arrival of higher ranking individual B can cause A to exit the feeding area hastily to allow B to occupy the location. Here B displaces A. Attacking could be inferred from an initial sudden rise in velocity of an individual A in the direction of another individual B. These behaviors can then be combined to recover the directed graph representing the social network of the animals.

Currently we quantify the dominance relationship $TA_{ij}$ using displacements and withdrawals between monkeys $i$ and $j$. If $TA_{ij} > TA_{ji}$ then monkey $j$ dominates monkey $i$. To obtain $TA_{ij}$ we calculate the bearing $B_{ij}$ of monkey $i$ to monkey $j$ and $i$'s velocity as $V_i$. $DV_{ij}$ is the component of monkey $i$'s motion directed towards monkey $j$. It is given by the magnitude of the projection of $V_i$ onto the unit vector $B_{ij}$.

$$DV_{ij} = V_i \cdot B_{ij} \quad (1)$$

$DV_{ij}$ is a measure of monkey $i$s motion with respect to monkey $j$, a value in $[-1, 1]$. Negative values of $DV_{ij}$ indicate monkey $i$ is moving away from monkey $j$ whereas positive values indicate monkey $i$ is moving towards monkey $j$. Our events of interest are when $-1 \leq DV_{ij} \leq -0.7$ i.e. monkey $i$ is moving directly away or at a slight angle from monkey $j$. $TA_{ij}$ is the total number of all events where

$-1 \leq DV_{ij} \leq -0.7$. The hierarchy Matrix $H$ is given by

$$[H]_{ij} = \begin{cases} 1 & \text{if } TA_{ij} < TA_{ji} \text{ ( j dominates i because i moves away from j more often )}; \\ 0 & \text{if } TA_{ij} \geq TA_{ji} \text{ ( i dominates j )}. \end{cases} \quad (2)$$

See Figure 2 for the Hierarchy graph derived from matrix $H$.

## 3. CONCLUSION

The goal of this research is to automatically infer group social structure using behaviors that can be captured by only observing position data. In our future work we aim to incorporate more complex behaviors within a general probabilistic framework to improve the inferred social structures.

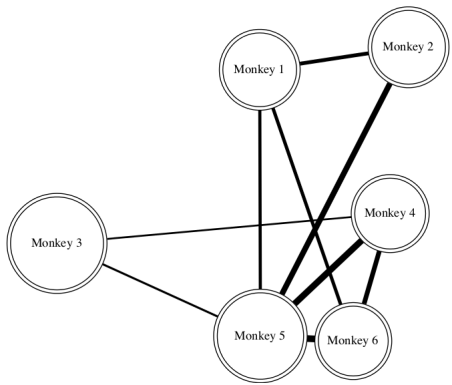
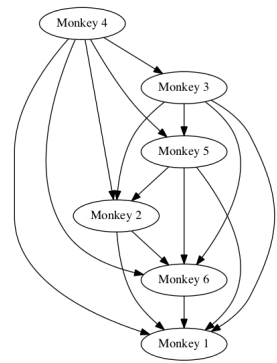

Fig. 1. An Undirected Affiliation Graph derived from the interaction matrix inferred from our dataset

Fig. 2. A Directed Hierarchy Graph representing the Hierarchy Matrix inferred from our dataset. The presence of an edge from Monkey 4 to Monkey 3 indicates that Monkey 4 dominates Monkey 3.

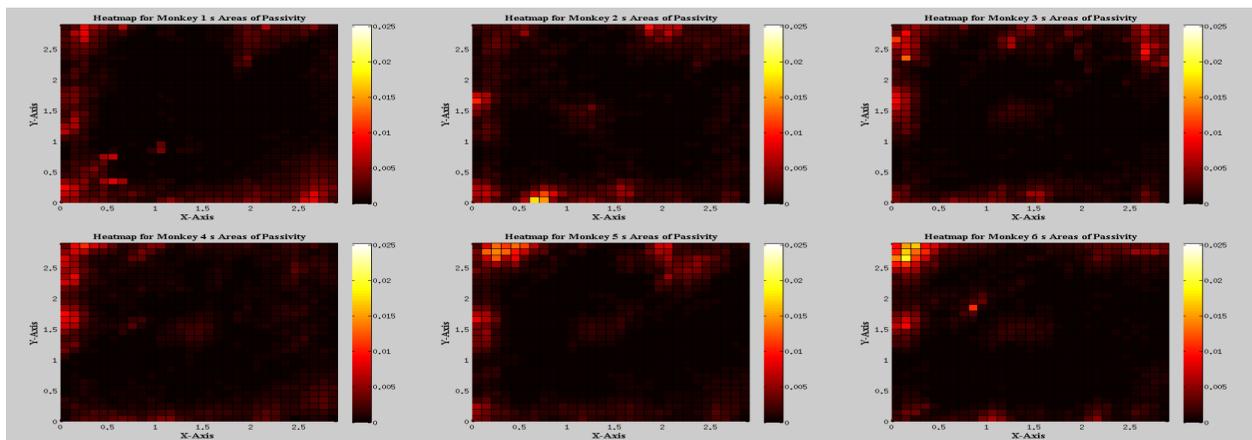

Fig. 3. Individual Heat Maps of passive activity for monkeys. Clockwise from top left : Heat Maps for monkey 1,2,3,6,5, and 4 Brighter areas indicate regions of the enclosure where monkeys remain stationary with greater frequency. The area of the enclosure (3m x 3m) has been divided into a 30 x 30 grid.